\documentclass[conference]{IEEEtran}
\IEEEoverridecommandlockouts
\usepackage{cite}
\usepackage{url}
\usepackage{amsmath,amssymb,amsfonts}
\usepackage{algorithmic}
\usepackage{graphicx}
\usepackage{textcomp}
\usepackage{xcolor}
\usepackage{listings}
\usepackage[numbers]{natbib}
\def\BibTeX{{\rm B\kern-.05em{\sc i\kern-.025em b}\kern-.08em
    T\kern-.1667em\lower.7ex\hbox{E}\kern-.125emX}}
    
\lstdefinestyle{mystyle}{
    backgroundcolor=\color{gray!10},
    basicstyle=\ttfamily\small,
    keywordstyle=\color{blue},
    commentstyle=\color{green!50!black},
    stringstyle=\color{red},
    emphstyle=\itshape\color{purple},
    showstringspaces=false,
    breaklines=true,
    frame=single,
    framexleftmargin=5mm,
    xleftmargin=5mm,
    linewidth=\linewidth
}

\begin{document}

\title{DevBots can co-design APIs}
\author{\IEEEauthorblockN{Vinicius Soares Silva Marques}
\IEEEauthorblockA{\textit{Department of Computer Science} \\
\textit{University of Brasília (UnB)}\\
Brasília, Brazil \\
marques.vinicius@gmail.com}
}

\maketitle

\begin{abstract}
DevBots are automated tools that perform various tasks in order to support software development. They are a growing trend and have been used in repositories to automate repetitive tasks, as code generators, and as collaborators in eliciting requirements and defining architectures. In this study, we analyzed 24 articles to investigate the state of the art of using DevBots in software development, trying to understand their characteristics, identify use cases, learn the relationship between DevBots and conversational software development, and discuss how prompt engineering can enable collaboration between human developers and bots. Additionally, we identified a gap to address by applying prompt engineering to collaborative API design between human designers and DevBots and proposed an experiment to assess what approach, between using Retrieval Augmented Generation or not, is more suitable. Our conclusion is that DevBots can collaborate with human API designers, but the two approaches have advantages and disadvantages.
\end{abstract}

\begin{IEEEkeywords}
DevBots, Conversational Software Development, API Design, Artificial Intelligence, Large Language Models, Generative Pre-Trained Transformers, Retrieval Augmented Generation
\end{IEEEkeywords}

\section{Introduction}
The use of bots to support software development activities has been a growing trend in recent years. Once restricted to automating repetitive tasks, such as commits to code repositories and opening and monitoring issues \cite{bi2023bothawk}, \cite{golzadeh2021evaluating}, today they are also used for code review \cite{copche2023can}, \cite{moguel2022bots}, creating test scripts \cite{kulkarni2021intelligent}, \cite{liao2023bdgoa}, and monitoring events \cite{copche2023can}. With the advance of Artificial Intelligence (AI), and especially with the popularization of Large Language Models (LLMs), they were also used for data extraction and analysis \cite{copche2023can}, \cite{moguel2022bots}, task recommendations \cite{copche2023can}, \cite{melo2023devbot} and collaborative development through code generation and requirements elicitation \cite{ahmad2023humanbot}.

Its use in repositories such as GitHub is widespread. Whether it's to automatically classify, comment on, monitor and close issues \cite{bi2023bothawk}, \cite{golzadeh2021evaluating}, \cite{golzadeh2021ground}, \cite{moguel2022bots}, or to automate commits \cite{dey2020detecting} \cite{golzadeh2021evaluating}, its presence is noted in many projects. Its use has evolved to Continuous Integration / Continuous Delivery (CI/CD) tracks \cite{bi2023bothawk}, \cite{copche2023can}, \cite{moguel2022bots}, with commits, dependency satisfaction and automated deploys, as well as event detection and monitoring \cite{copche2023can}.

It is also possible to recognize the presence of DevBots in code-related activities. Activities such as code review \cite{bi2023bothawk}, \cite{copche2023can}, \cite{moguel2022bots}, identifying and correcting bugs \cite{copche2023can}, \cite{moguel2022bots} or violations of coding standards \cite{moguel2022bots}, and static analysis \cite{copche2023can}, \cite{moguel2022bots} benefit from the use of bots in terms of productivity \cite{moguel2022bots}, \cite{erlenhov2020bots}, \cite{kulkarni2021intelligent}, \cite{nembhard2023teaming}. With the popularization of LLMs, more complex activities are now being carried out by or in collaboration with DevBots. Extracting, analyzing and sharing data is one of them. In addition, bots can recommend tasks and improvements \cite{copche2023can}, \cite{moguel2022bots}. More recently, conversational software development has allowed for the elicitation of requirements, collaborative design of architectures \cite{ahmad2023humanbot} and the generation of code, specifications and documentation \cite{kulkarni2021intelligent}, \cite{Parashar2022revolutionary}, bringing numerous opportunities for future work.

This study selected and analyzed 24 articles that made it possible to investigate the state of the art of using DevBots in software development. Their main characteristics and use cases were identified. It was also possible to reveal the relationship between DevBots and conversational software development. In addition, the study sought to discuss how prompt engineering can enable collaboration between human developers and bots. Finally, we created and conducted an experiment in which we created API specifications with the aid of an LLM and assessed whether a Retrieval Augmented Generation approach helps in the process.

\section{State of the Art}

\subsection{Methodology}

A Systematic Literature Review (SLR) was carried out following the guidelines of \citet{kitchenham2009slr} to investigate the use of bots in software development. To do this, research questions were created, a search string was built, databases were selected, selection criteria were defined, a quality assessment was carried out and finally the data was extracted and analyzed.

\subsection{Research Questions}
In this work, the aim is to reveal the state of the art in the use of bots in software development, to enable the identification of possible gaps for future studies. The research questions are as follows (Table \ref{table:research_questions}):

\begin{table}[ht]
\centering
\begin{tabular}{|c|p{6cm}|} 
\hline
\textbf{RQ} & \textbf{Research Questions} \\
\hline
RQ.1 & What are the most common features of DevBots? \\
\hline
RQ.2 & What are some use cases of DevBots? \\
\hline
RQ.3 & What is the relationship between DevBots and conversational software development? \\
\hline
RQ.4 & How does prompt engineering enable the use of DevBots in software design and development? \\
\hline
\end{tabular}
\caption{List of Research Questions}
\label{table:research_questions}
\end{table}

\subsection{Search String}\label{AA}
As the two main subjects to be investigated in this work cover DevBots and conversational software development, we used these keywords to construct the search string:

\begin{lstlisting}[style=mystyle]
"conversational software development" OR devbots
\end{lstlisting}

\subsection{Databases}
The databases selected were ACM Digital Library, IEEE Xplore, Scopus, Web of Science and, to ensure that the most recent results were included, arXiv. To augment the search, Google Scholar was also included, with the following search string:

\begin{lstlisting}[style=mystyle]
"AI-assisted software development"
\end{lstlisting}

\subsection{Selection Criteria}
Inclusion and exclusion criteria were defined. The inclusion criteria concern the topics covered by the study, the period and the type of articles accepted in the context of this study. The inclusion criteria are listed below:

\begin{enumerate}
    \item Papers which describe a use case or the development of a DevBot, or literature reviews on the subjects.
    \item Published works (including preprints) from 2020 onwards (year of publication of GTP-3 \cite{brown2020language}).
\end{enumerate}

The exclusion criteria aim to remove articles that do not meet the inclusion criteria and do not contribute to the objectives of this study. The exclusion criteria are shown below:

\begin{enumerate}
    \item Duplicates between the databases.
    \item Use of search terms in contexts other than the context of this work, or unrelated to the main subjects of this study (DevBots and conversational software development).
    \item Compendia of works (as opposed to unique works), PhD or Masters thesis, industry papers and abstract only works.
    \item Papers not written in English.
\end{enumerate}

\subsection{Quality Assessment}
After applying the inclusion criteria, titles and abstracts had to be analyzed in order to apply the exclusion criteria and ensure that the papers selected adhered to the objectives of this study. Papers in which the search terms did not appear in the title or abstract were discarded, as were those in which the terms appeared in contexts other than those proposed in this article. Results that referred to compendia or collections of works were also discarded.

\subsection{Conducting}
\begin{figure}[ht]
  \centering
  \includegraphics[width=1.0\linewidth]{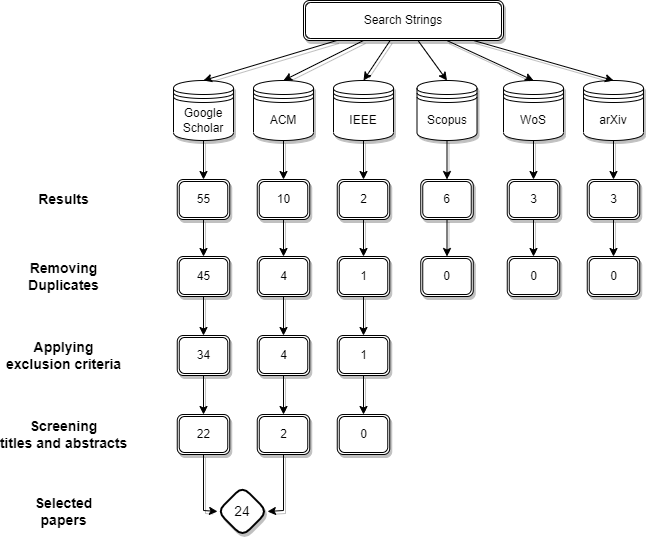}
  \caption{Selection of papers}
  \label{fig:papers}
\end{figure}

Initially, the searches returned a total of 79 results based on the search strings and time interval defined. We then applied the criterion of excluding duplicates, which reduced this number to 60, with only 5 not appearing in either Google Scholar or other databases (4 from ACM DL and 1 from IEEE Xplore). 11 articles not written in English and 13 papers in an inappropriate format (collections of papers, doctoral or master's theses, industry articles and abstracts) were excluded. The remaining papers were screened for titles and abstracts and finally 24 articles were selected for data extraction and analysis (Figure \ref{fig:papers}).

\subsection{Data Extraction}
The data was extracted considering the objectives of this study and the research questions. In this way, each paper was analyzed to extract the most common characteristics of DevBots (RQ.1); the most common use cases for DevBots (RQ.2); the relationships between DevBots and conversational software development (RQ.3); and insights into how prompt engineering can enable the use of bots in software development (RQ.4).

\section{Results of the SLR}
\subsection{Related Works}
\citet{necula2023artificial} gave an insight into the impact of AI on the software engineering profession, highlighting the new challenges posed by the growing adoption of AI in the field of Software Engineering and the skills that will be demanded of professionals.

\citet{wu2022bots}, \citet{melo2023devbot}, \citet{wyrich2021bots}, \citet{wessel2022software}, \citet{moguel2022bots}, \citet{golzadeh2021ground} and \citet{bi2023bothawk} investigated the use of DevBots on social development platforms such as GitHub.

\citet{tony2022devbots} and \citet{santhanam2022bots} sought to understand how developers relate to conversational software development supported by DevBots. \citet{pothukuchi2023impact}, \citet{Parashar2022revolutionary} and \citet{kulkarni2021intelligent} discussed how to use AI throughout the software development lifecycle.

\citet{erlenhov2020bots} worked on defining the DevBot concept, usage scenarios and challenges.

\subsection{DevBots Features (RQ.1)}
To answer RQ.1, the papers were analyzed for characteristics commonly associated with DevBots, whether perceived or desired. The most cited feature was \textbf{the ability to interact with the developer through natural language processing}, noted by \citet{ahmad2023humanbot}, \citet{erlenhov2020bots},  \citet{kulkarni2021intelligent}, \citet{nembhard2023teaming}, \citet{Parashar2022revolutionary}, \citet{pothukuchi2023impact}, \citet{santhanam2022bots} and \citet{tony2022devbots}. This was followed by features related to this capability, such as \textbf{AI-driven collaboration}, cited by \citet{ahmad2023humanbot}, \citet{kulkarni2021intelligent}, \citet{nembhard2023teaming} and \citet{Parashar2022revolutionary}; \textbf{intelligent behavior}, noted by \citet{ahmad2023humanbot}, \citet{erlenhov2020bots}, \citet{kulkarni2021intelligent} and \citet{Parashar2022revolutionary}; \textbf{the ability to answer questions}, mentioned by \citet{ahmad2023humanbot}, \citet{nembhard2023teaming}, \citet{tony2022devbots} and \citet{melo2023devbot}; and \textbf{interaction with the user}, informed by  \citet{erlenhov2020challenges}, \citet{nembhard2023teaming}, \citet{santhanam2022bots} and \citet{tony2022devbots}. \textbf{Presenting itself as a conversational agent} was noted by \citet{ahmad2023humanbot} and \citet{erlenhov2020bots}. \textbf{Creating synergy between human collaborators and bots} was a feature cited by \citet{ahmad2023humanbot} and \citet{nembhard2023teaming}.

Characteristics linked to \textbf{the ability to provide automation and autonomous behavior} were also cited, like \textbf{the ability to generate code, specifications and documentation} was cited by \citet{ahmad2023humanbot}, \citet{kulkarni2021intelligent} and \citet{nembhard2023teaming}. \textbf{The ability to build knowledge base systems} was a feature cited by \citet{kulkarni2021intelligent}, \citet{Parashar2022revolutionary} and \citet{tony2022devbots}. \textbf{The ability to make recommendations} was noted by \citet{ahmad2023humanbot} and \citet{tony2022devbots}.

\citet{erlenhov2020bots} cites the following characteristics of DevBots: \textbf{being triggered agents, presenting identity, being scalable and presenting adaptability}. This last characteristic was also cited by \citet{tony2022devbots}, as well as \textbf{clarity in interactions, reliability, accessibility, readiness and support for integration with the existing development environment}. \citet{uzair2023six} cites \textbf{modularity, reusability and maintainability}.

\subsection{Use Cases (RQ.2)}
The findings that answer RQ.2 are summarized in Table \ref{table:use_cases}. It is important to note that although the automation of repetitive tasks is among the most cited, more complex use cases are presented by several authors.

\begin{table*}[t]
  \centering
  \begin{tabular}{|p{4cm}|p{10cm}|c|}
    \hline
    Use case & Paper & \# of papers \\
    \hline
    Issue and pull request lifecycle management & \citet{bi2023bothawk}, \citet{golzadeh2021evaluating}, \citet{golzadeh2021ground}, \citet{moguel2022bots}, \citet{erlenhov2020bots}, \citet{liao2023bdgoa}, \citet{wyrich2021bots}, \citet{wu2022bots}, \citet{melo2023devbot}
 & 9 \\
    \hline
    Code review & \citet{bi2023bothawk}, \citet{copche2023can}, \citet{moguel2022bots}, \citet{erlenhov2020bots}, \citet{liao2023bdgoa}, \citet{nembhard2023teaming}, \citet{wessel2022software}, \citet{uzair2023six}, \citet{wu2022bots} & 9 \\
    \hline
    Testing & \citet{copche2023can}, \citet{erlenhov2020challenges}, \citet{kulkarni2021intelligent}, \citet{liao2023bdgoa}, \citet{Parashar2022revolutionary}, \citet{pothukuchi2023impact}, \citet{wessel2022software}, \citet{necula2023artificial}, \citet{wu2022bots} & 9 \\
    \hline
    Bug fixing and code debugging & \citet{copche2023can}, \citet{moguel2022bots}, \citet{erlenhov2020bots}, \citet{nembhard2023teaming}, \citet{pothukuchi2023impact}, \citet{santhanam2022bots,} \citet{tony2022devbots}, \citet{wessel2022software}, \citet{wyrich2021bots} & 9 \\
    \hline
    CI/CD & \citet{bi2023bothawk}, \citet{copche2023can}, \citet{moguel2022bots}, \citet{erlenhov2020bots}, \citet{pothukuchi2023impact}, \citet{wu2022bots} & 6 \\
    \hline
    Code analysis and verification & \citet{copche2023can}, \citet{moguel2022bots}, \citet{erlenhov2020bots}, \citet{nembhard2023teaming}, \citet{santhanam2022bots}, \citet{tony2022devbots} & 6 \\
    \hline
    Extracting and analysing data & \citet{copche2023can}, \citet{moguel2022bots}, \citet{erlenhov2020bots}, \citet{Parashar2022revolutionary}, \citet{wessel2022software} & 5 \\
    \hline
    Dependency management & \citet{erlenhov2022dependency}, \citet{liao2023bdgoa}, \citet{uzair2023six}, \citet{wyrich2021bots}, \citet{wu2022bots} & 5 \\
    \hline
    Coding and implementing software & \citet{kulkarni2021intelligent}, \citet{Parashar2022revolutionary}, \citet{pothukuchi2023impact}, \citet{santhanam2022bots}, \citet{uzair2023six} & 5 \\
    \hline
    Software requirements specification & \citet{ahmad2023humanbot}, \citet{kulkarni2021intelligent}, \citet{Parashar2022revolutionary}, \citet{pothukuchi2023impact}, \citet{uzair2023six} & 5 \\
    \hline
    Automatic issue commenting & \citet{bi2023bothawk}, \citet{golzadeh2021evaluating}, \citet{golzadeh2021ground}, \citet{moguel2022bots} & 4 \\
    \hline
    Task recommendation & \citet{copche2023can}, \citet{moguel2022bots}, \citet{liao2023bdgoa}, \citet{melo2023devbot} & 4 \\
    \hline
    Software design & \citet{ahmad2023humanbot}, \citet{kulkarni2021intelligent}, \citet{Parashar2022revolutionary}, \citet{pothukuchi2023impact} & 4 \\
    \hline
    Bug detection & \citet{moguel2022bots}, \citet{pothukuchi2023impact}, \citet{wessel2022software}, \citet{uzair2023six} & 4 \\
    \hline
    Code refactoring & \citet{liao2023bdgoa}, \citet{santhanam2022bots}, \citet{wessel2022software}, \citet{wyrich2021bots} & 4 \\
    \hline
    Detecting and mitigating vulnerabilities & \citet{nembhard2023teaming}, \citet{Parashar2022revolutionary}, \citet{pothukuchi2023impact}, \citet{tony2022devbots} & 4 \\
    \hline
    Detecting and monitoring events & \citet{copche2023can}, \citet{liao2023bdgoa}, \citet{wu2022bots} & 3 \\
    \hline
    Connecting with stakeholders & \citet{copche2023can}, \citet{Parashar2022revolutionary}, \citet{pothukuchi2023impact} & 3 \\
    \hline
    Providing feedback & \citet{copche2023can}, \citet{moguel2022bots}, \citet{wessel2022software} & 3 \\
    \hline
    Software engineering teaching & \citet{kulkarni2021intelligent}, \citet{wessel2022software}, \citet{necula2023artificial} & 3 \\
    \hline
    Prototyping & \citet{Parashar2022revolutionary}, \citet{pothukuchi2023impact}, \citet{santhanam2022bots} & 3 \\
    \hline
    Sharing information & \citet{copche2023can}, \citet{moguel2022bots} & 2 \\
    \hline
    Commiting to repositories & \citet{golzadeh2021evaluating}, \citet{dey2020detecting} & 2 \\
    \hline
    Violation of coding standards detection & \citet{moguel2022bots}, \citet{wu2022bots} & 2 \\
    \hline
    24/7 task handling & \citet{erlenhov2020bots}, \citet{wessel2022software} & 2 \\
    \hline
    Integrating newcomers into the team & \citet{liao2023bdgoa}, \citet{wessel2022software} & 2 \\
    \hline
    Commenting in social media & \citet{copche2023can} & 1 \\
    \hline
    Build error identification & \citet{moguel2022bots} & 1 \\
    \hline
  \end{tabular}
  \caption{DevBot Use Cases}
  \label{table:use_cases}
\end{table*}

\subsection{Relationships between DevBots and conversational software development (RQ.3)}
The answers to RQ.3 are not simple. What can be inferred from the literature is that the use of DevBots for conversational software development is still in its infancy, although we can already see initiatives in this regard. According to \citet{ahmad2023humanbot}, \citet{Parashar2022revolutionary}, \citet{pothukuchi2023impact} and \citet{uzair2023six}, \textbf{the translation of high-level requirements into architectural models and evaluation scenarios is a relationship that can emerge from the use of AI in software development}, taking advantage of conversational strategies. To do this, according to \citet{erlenhov2020bots}, \citet{nembhard2023teaming} and \citet{tony2022devbots}, \textbf{these technologies need to be able to parse intents}, to infer what the developer wants and present a relevant solution.

\subsection{DevBot Usage Enablement by Prompt Engineering (RQ. 4)}
According to \citet{ahmad2023humanbot}, \citet{pothukuchi2023impact} and \citet{uzair2023six}, \textbf{to enable the use of DevBots through Prompt Engineering the bots need to be able to create the initial specification and iterative enhancements from carefully prepared prompts, requiring human supervision for the generation of code, specifications and documentation}. \citet{ahmad2023humanbot}, \citet{moguel2022bots}, \citet{pothukuchi2023impact} and \citet{uzair2023six} point out that these scenarios require the \textbf{extraction of requirements based on conversations}. Another possible relationship, according to \citet{ahmad2023humanbot} and \citet{tony2022devbots}, could be in \textbf{obtaining testing and debugging scenarios from conversational interactions between humans and bots}.

\section{Discussion of the SLR}
Although many authors have applied AI to software development, there is still room for future research. In addition to applications in repetitive task automation \cite{bi2023bothawk}, \cite{golzadeh2021evaluating}, \cite{dey2020detecting}, \cite{golzadeh2021ground}, \cite{liao2023bdgoa}, some studies have employed DevBots to actually collaborate in more complex work. \citet{nembhard2023teaming} developed a framework to integrate virtual assistants with human developers through verbal conversations to discuss security aspects and produce more secure software.

\citet{ahmad2023humanbot} described a case study involving collaboration between a novice software architect and ChatGPT for software architectural analysis, synthesis, and evaluation using architecture-centric software engineering (ACSE) concepts, and concludes that ChatGPT can simulate the role of an architect to support or even lead ACSE under human oversight and support for collaborative architecture.

We are aware of other studies that were not included in our SLR. \citet{blocklove2023} performed a case study on the use of ChatGPT for hardware design and programming. Through a ChatGPT prompt conversation, a design for an 8-bit shift register is generated (along with a test bench) and evaluated and tested for errors, first by means of a simulation tool, then with the help of human feedback at three levels (simple, moderate, and advanced). The findings show that the approach is useful as long as it is used for co-design in collaboration with a human designer, since ChatGPT cannot generate a complete design with only the initial human interaction.

\citet{Gupta2022} provided an AI-augmented framework for persona pool creation, software requirements specification, and usability evaluation that can be used in the requirements analysis phase to improve the usability of software under design. This framework proposes ontologies for representing elemental and abstracted knowledge that can be employed to maximize the effectiveness of AI techniques applied to usability evaluation.

However, there are still gaps in the use of AI in collaboration with humans. Our study aimed to use conversational design to perform API design. To achieve this, we proposed an experiment in which GPT-3.5 and a human API designer create the collaborative design of an API and its OpenAPI Specification (OAS) through a single prompt using Prompt Engineering (PE) \cite{Liu2023} and Prompt Optimization (PO) \cite{yang2023large} techniques, and evaluate whether Retrieval Augmented Generation (RAG) \cite{chen2023benchmarking} leads to better results.

\section{Experiment design}
The experiment consisted of a series of steps executed in Microsoft Azure Machine Learning Studio. First, we created a prompt to request for the creation of an OAS. This prompt was executed in two pipelines: one containing an RAG structure using the text-embedding-ada-002 model together with a query to the gpt-35-turbo-0613 model and the other only with a direct query to the gpt-35-turbo-0613 model. The prompts for the two pipelines were crafted slightly differently, with the only difference being that the one dedicated to the pipeline with RAG contained two questions to reinforce the discovery, with the help of the embedded documents, of the server URL for the three environments (development, homologation, and production) and the OAuth scopes within the security schema. The pipeline with RAG was fed (through embedding and vector indexing) with PDF files produced from the OAS publicly available on Banco do Brasil's developer portal \footnote{\url{https://www.bb.com.br/site/developers/solucoes-api-bb/}}, so that it would be possible to assess the impact of the presence of OAS models considered correct as an input for the generation of OAS by LLM.

This was performed by alternately triggering the two pipelines 10 times each. The response went through a cleaning process that consisted of replacing the characters \verb|\n| and \verb|\"| with a line break and a double quotation mark, respectively, and removing any text generated by the LLM in addition to the JSON containing the OAS. The JSON was then copied and pasted into Smartbear's Swagger Editor \footnote{\url{https://editor.swagger.io/}} to check the parameters for correctness. The replication package is available at GitHub \footnote{\url{https://github.com/marques-vinicius/OAGen}}.

\subsection{Correctness assessment}\label{subsection:correctness_assessment}
The following correctness parameters were selected:
\begin{enumerate}
    \item \textbf{No syntax errors:} OAS free of errors, in which case the attempt received a score of 0.2 in this regard, otherwise it scored 0.00;
    \item \textbf{Renders correctly:} OAS presented errors that prevented it from being rendered, in which case the attempt received a score of -0.1. If the error was just a missing character that, when inserted, allowed rendering, or if the errors did not prevent rendering, it received a score of 0.00 in this regard;
    \item \textbf{Paths according to the REST standard \cite{fielding2000architectural}:} in which case the attempt received a score of 0.2 in this regard, otherwise it scored 0.00;
    \item \textbf{Methods according to the REST standard}: in which case the attempt received a score of 0.15 in this regard, otherwise it scored 0.00;
    \item \textbf{Functional security scheme:} OAS presented a correct security scheme with two OAuth scopes (one for queries and one for other requests) and the scopes were correctly linked, in which case the attempt received a score of 0.2 in this regard, otherwise it scored 0.00;
    \item \textbf{Examples of requests included:} in which case the attempt received a score of 0.05 in this regard, otherwise it scored 0.00;
    \item \textbf{Examples of successful responses included:} if both endpoints had successful response examples (with status code 200 for GET and 201 for POST), the attempt received a score of 0.05, but if only one of them had a response example, the attempt received a score of 0.03, otherwise it scored 0.00;
    \item \textbf{Error status codes included:} at least one status code besides the success code was included, in which case the attempt received a score of 0.05 in this regard, otherwise it scored 0.00;
    \item \textbf{One server for each environment:} a server URL was designated for each of the three environments (development, homologation and production), in which case the attempt received a score of 0.1 in this regard, otherwise it scored 0.00.
\end{enumerate}
These parameters were scored such that the attempt received a final score of 1 if all the requirements were met, and 0 if none were met. All OAS provided as benchmarks to the RAG pipeline scored 1 according to these parameters.

\section{Results and discussion of the experiment}
The scores obtained for each attempt are presented in Table \ref{tab:table_correctness_assessment}. Greater consistency can be observed in the OASs generated by the RAG pipeline, although this does not necessarily translate into a lack of syntax errors. None of the RAG pipeline attempts scored above 0.80, although the pipeline without RAG obtained a score of 0.88 and surpassed the average of the other pipeline in three attempts.

\begin{table*}[t]
    \centering
    \scriptsize
    \begin{tabular}{|c|*{10}{r|}*{10}{r|}}
        \hline
        \multicolumn{1}{|c|}{} & \multicolumn{10}{c|}{Without RAG} & \multicolumn{10}{c|}{With RAG} \\
        \hline
        \# & 1 & 2 & 3 & 4 & 5 & 6 & 7 & 8 & 9 & 10 & 1 & 2 & 3 & 4 & 5 & 6 & 7 & 8 & 9 & 10 \\
        \hline
        1 & 0.00 & 0.00 & 0.20 & 0.00 & 0.20 & 0.00 & 0.00 & 0.00 & 0.20 & 0.00 & 0.00 & 0.00 & 0.00 & 0.00 & 0.00 & 0.00 & 0.00 & 0.00 & 0.00 & 0.00 \\
        \hline
        2 & 0.00 & 0.00 & 0.00 & 0.00 & 0.00 & 0.00 & 0.00 & 0.00 & 0.00 & 0.00 & 0.00 & 0.00 & 0.00 & 0.00 & 0.00 & 0.00 & 0.00 & 0.00 & 0.00 & 0.00 \\
        \hline
        3 & 0.00 & 0.00 & 0.20 & 0.00 & 0.20 & 0.00 & 0.00 & 0.00 & 0.20 & 0.00 & 0.20 & 0.00 & 0.20 & 0.20 & 0.20 & 0.00 & 0.00 & 0.20 & 0.20 & 0.20 \\
        \hline
        4 & 0.15 & 0.15 & 0.15 & 0.15 & 0.15 & 0.15 & 0.15 & 0.15 & 0.15 & 0.15 & 0.15 & 0.15 & 0.15 & 0.15 & 0.15 & 0.15 & 0.15 & 0.15 & 0.15 & 0.15 \\
        \hline
        5 & 0.20 & 0.00 & 0.20 & 0.00 & 0.00 & 0.20 & 0.20 & 0.20 & 0.00 & 0.00 & 0.20 & 0.20 & 0.20 & 0.20 & 0.20 & 0.20 & 0.20 & 0.20 & 0.20 & 0.20 \\
        \hline
        6 & 0.05 & 0.05 & 0.05 & 0.05 & 0.05 & 0.05 & 0.05 & 0.05 & 0.05 & 0.05 & 0.05 & 0.05 & 0.05 & 0.05 & 0.05 & 0.05 & 0.05 & 0.05 & 0.05 & 0.05 \\
        \hline
        7 & 0.05 & 0.03 & 0.03 & 0.03 & 0.03 & 0.03 & 0.05 & 0.00 & 0.00 & 0.03 & 0.03 & 0.03 & 0.03 & 0.03 & 0.03 & 0.03 & 0.03 & 0.03 & 0.03 & 0.03 \\
        \hline
        8 & 0.05 & 0.05 & 0.05 & 0.05 & 0.05 & 0.05 & 0.05 & 0.05 & 0.05 & 0.05 & 0.05 & 0.05 & 0.05 & 0.05 & 0.05 & 0.05 & 0.05 & 0.05 & 0.05 & 0.05 \\
        \hline
        9 & 0.10 & 0.10 & 0.00 & 0.10 & 0.00 & 0.10 & 0.10 & 0.10 & 0.00 & 0.00 & 0.10 & 0.10 & 0.10 & 0.00 & 0.10 & 0.10 & 0.10 & 0.10 & 0.10 & 0.10 \\
        \hline
        $\Sigma$ & 0.60 & 0.38 & 0.88 & 0.38 & 0.68 & 0.78 & 0.60 & 0.75 & 0.45 & 0.28 & 0.78 & 0.58 & 0.78 & 0.68 & 0.78 & 0.58 & 0.58 & 0.78 & 0.78 & 0.78 \\
        \hline
        \multicolumn{1}{|c|}{$\overline{\Sigma}$} & \multicolumn{10}{c|}{0.578} & \multicolumn{10}{c|}{0.71} \\
        \hline
    \end{tabular}
    \caption{This table shows the correctness parameters presented in Subsection \ref{subsection:correctness_assessment}, represented by the numbers 1 to 9 below the \# symbol; the attempts without and with RAG, represented by the numbers 1 to 10 to the right of the \# symbol; the scores for each attempt, represented by the scores below each atttempt; and the average of the sum of the scores for all attempts ($\overline{\Sigma}$), present in the last two cells of the last row of the table.}
    \label{tab:table_correctness_assessment}
\end{table*}

The most common errors in OAS generation were syntax errors (60\% of the attempts without RAG and 100\% of the attempts with RAG), the absence of a functional security scheme (50\% of the attempts without RAG, while the pipeline with RAG was correct in all attempts), the creation of paths outside the REST standard (70\% of attempts by the pipeline without RAG and 30\% of the attempts with RAG), and the lack of one or more servers (40\% of attempts by the pipeline without RAG, whereas the pipeline with RAG generated this error only once). The absence of response examples for at least the success status code (200 or 201) was observed in both pipelines (80\% of the attempts by the pipeline without RAG and 100\% of the attempts by the RAG pipeline). Both pipelines correctly chose the methods for each endpoint in every attempt, although this may be due to the fact that only one query (GET) and one resource creation (POST) were requested, without the possibility of using the other available methods (PUT, PATCH, DELETE, OPTIONS, HEAD), which may have made the choice easier. In only one case (in the pipeline without RAG) was an OAS generated with an error that prevented rendering. This error consisted of the omission of a ':' character between the name of a key and its value. By including this character, the problem was solved, and the OAS was rendered correctly. In two attempts, the pipeline without RAG correctly generated the OAuth scopes (attempts 5 and 9), although it made mistakes in linking the scopes to the endpoints and thus the attempt was considered wrong. In one case, the pipeline without RAG, although it generated more than one status code, was limited to including only 200 and 400 (attempt 5, and in the others, it generated at least three different status codes). In three attempts (2, 7 and 9), the pipeline without RAG introduced errors beyond those related to the security scheme, and in one (attempt 7), these were the only errors made (in the others, the security scheme was not generated correctly either). As for the RAG pipeline, all attempts had syntax errors, none of them related to the security scheme.

It can be seen that the results of the RAG pipeline were much more consistent with an overall variance of only 0.008 compared to 0.035 for the pipeline without RAG, but this did not translate into a better overall quality for OAS, as there were plenty of syntax errors in all attempts with RAG. However, the average score for all attempts was significantly better than that for the pipeline without RAG. In contrast, three attempts in the pipeline without RAG outperformed the average of the attempts in the pipeline with RAG, suggesting that consistency is the only real advantage of using RAG to generate OAS, but further studies may be performed with a much larger number of attempts for validation. On the other hand, in the third attempt, the pipeline without RAG obtained a score very close to the maximum (0.88) and above all attempts with RAG, as well as another two attempts with a score higher than the average of the other pipeline. In short, to determine which approach produces the best results with a greater degree of assertiveness, it would be necessary to carry out a much larger number of trials. Therefore, we consider the results to be inconclusive regarding the impact of RAG on LLM performance in generating OAS.

\section{Conclusions}
DevBots are widely used in various activities related to software development; however, to the best of our knowledge, there are no records of their use in the collaborative design of APIs with humans. To address this gap, we created and conducted an experiment using an LLM, GPT-3.5, to create the OAS of an API with and without the help of an RAG approach. Our findings suggest that it is possible to use DevBots in collaboration with a human API designer, but the results are inconclusive regarding the use of RAG for this purpose. Future studies could evaluate this approach better by repeating the experiment with significantly more trials.

\bibliographystyle{unsrtnat}
\bibliography{main}

\begin{thebibliography}{31}
\providecommand{\natexlab}[1]{#1}
\providecommand{\url}[1]{\texttt{#1}}
\expandafter\ifx\csname urlstyle\endcsname\relax
  \providecommand{\doi}[1]{doi: #1}\else
  \providecommand{\doi}{doi: \begingroup \urlstyle{rm}\Url}\fi

\bibitem[Bi et~al.(2023)Bi, Zhu, Wang, Xia, Khan, and Pu]{bi2023bothawk}
Fenglin Bi, Zhiwei Zhu, Wei Wang, Xiaoya Xia, Hassan~Ali Khan, and Peng Pu.
\newblock Bothawk: An approach for bots detection in open source software projects.
\newblock \emph{arXiv preprint arXiv:2307.13386}, 2023.

\bibitem[Golzadeh et~al.(2021{\natexlab{a}})Golzadeh, Decan, and Mens]{golzadeh2021evaluating}
Mehdi Golzadeh, Alexandre Decan, and Tom Mens.
\newblock Evaluating a bot detection model on git commit messages.
\newblock \emph{arXiv preprint arXiv:2103.11779}, 2021{\natexlab{a}}.

\bibitem[Copche et~al.(2023)Copche, Pessanha, Durelli, Eler, and Endo]{copche2023can}
Rubens Copche, Yohan~Duarte Pessanha, Vinicius Durelli, Marcelo~Medeiros Eler, and Andre~Takeshi Endo.
\newblock Can a chatbot support exploratory software testing? preliminary results.
\newblock \emph{arXiv preprint arXiv:2307.05807}, 2023.

\bibitem[Moguel-S\'anchez et~al.(2022)Moguel-S\'anchez, Mart\'inez-Palacios, Ochar\'an-Hern\'andez, Lim\'on, and S\'anchez-Garc\'ia]{moguel2022bots}
Ricardo Moguel-S\'anchez, C\'esar~Sergio Mart\'inez-Palacios, Jorge~Octavio Ochar\'an-Hern\'andez, Xavier Lim\'on, and \'Angel~J. S\'anchez-Garc\'ia.
\newblock Bots and their uses in software development: A systematic mapping study.
\newblock In \emph{2022 10th International Conference in Software Engineering Research and Innovation (CONISOFT)}, pages 140--149, Oct 2022.
\newblock \doi{10.1109/CONISOFT55708.2022.00027}.

\bibitem[Kulkarni et~al.(2022)Kulkarni, Kolhe, and Kulkarni]{kulkarni2021intelligent}
Vaishnavi Kulkarni, Anurag Kolhe, and Jay Kulkarni.
\newblock Intelligent software engineering: The significance of artificial intelligence techniques in enhancing software development lifecycle processes.
\newblock In Ajith Abraham, Niketa Gandhi, Thomas Hanne, Tzung-Pei Hong, Tatiane Nogueira~Rios, and Weiping Ding, editors, \emph{Intelligent Systems Design and Applications}, pages 67--82, Cham, 2022. Springer International Publishing.
\newblock ISBN 978-3-030-96308-8.

\bibitem[Liao et~al.(2023)Liao, Huang, Zhang, Wu, and Cheng]{liao2023bdgoa}
Zhifang Liao, Xuechun Huang, Bolin Zhang, Jinsong Wu, and Yu~Cheng.
\newblock Bdgoa: A bot detection approach for github oauth apps.
\newblock \emph{Intelligent and Converged Networks}, 2023.
\newblock \doi{10.23919/ICN.2023.0006}.
\newblock URL \url{https://www.sciopen.com/article/10.23919/ICN.2023.0006}.

\bibitem[Melo(2023)]{melo2023devbot}
Glaucia Melo.
\newblock Designing adaptive developer-chatbot interactions: Context integration, experimental studies, and levels of automation.
\newblock In \emph{Proceedings of the 45th International Conference on Software Engineering: Companion Proceedings}, ICSE '23, page 235–239. IEEE Press, 2023.
\newblock ISBN 9798350322637.
\newblock \doi{10.1109/ICSE-Companion58688.2023.00064}.
\newblock URL \url{https://doi.org/10.1109/ICSE-Companion58688.2023.00064}.

\bibitem[Ahmad et~al.(2023)Ahmad, Waseem, Liang, Fahmideh, Aktar, and Mikkonen]{ahmad2023humanbot}
Aakash Ahmad, Muhammad Waseem, Peng Liang, Mahdi Fahmideh, Mst~Shamima Aktar, and Tommi Mikkonen.
\newblock Towards human-bot collaborative software architecting with chatgpt.
\newblock In \emph{Proceedings of the 27th International Conference on Evaluation and Assessment in Software Engineering}, EASE '23, page 279–285, New York, NY, USA, 2023. Association for Computing Machinery.
\newblock ISBN 9798400700446.
\newblock \doi{10.1145/3593434.3593468}.

\bibitem[Golzadeh et~al.(2021{\natexlab{b}})Golzadeh, Decan, Legay, and Mens]{golzadeh2021ground}
Mehdi Golzadeh, Alexandre Decan, Damien Legay, and Tom Mens.
\newblock A ground-truth dataset and classification model for detecting bots in github issue and pr comments.
\newblock \emph{Journal of Systems and Software}, 175:\penalty0 110911, 2021{\natexlab{b}}.

\bibitem[Dey et~al.(2020)Dey, Mousavi, Ponce, Fry, Vasilescu, Filippova, and Mockus]{dey2020detecting}
Tapajit Dey, Sara Mousavi, Eduardo Ponce, Tanner Fry, Bogdan Vasilescu, Anna Filippova, and Audris Mockus.
\newblock Detecting and characterizing bots that commit code.
\newblock In \emph{Proceedings of the 17th International Conference on Mining Software Repositories}, MSR '20, page 209–219, New York, NY, USA, 2020. Association for Computing Machinery.
\newblock ISBN 9781450375177.
\newblock \doi{10.1145/3379597.3387478}.
\newblock URL \url{https://doi.org/10.1145/3379597.3387478}.

\bibitem[Erlenhov et~al.(2020{\natexlab{a}})Erlenhov, Neto, and Leitner]{erlenhov2020bots}
Linda Erlenhov, Francisco Gomes de~Oliveira Neto, and Philipp Leitner.
\newblock An empirical study of bots in software development: Characteristics and challenges from a practitioner’s perspective.
\newblock In \emph{Proceedings of the 28th ACM Joint Meeting on European Software Engineering Conference and Symposium on the Foundations of Software Engineering}, ESEC/FSE 2020, page 445–455, New York, NY, USA, 2020{\natexlab{a}}. Association for Computing Machinery.
\newblock ISBN 9781450370431.
\newblock \doi{10.1145/3368089.3409680}.
\newblock URL \url{https://doi.org/10.1145/3368089.3409680}.

\bibitem[Nembhard and Carvalho(2023)]{nembhard2023teaming}
Fitzroy~D. Nembhard and Marco~M. Carvalho.
\newblock Teaming humans with virtual assistants to detect and mitigate vulnerabilities.
\newblock In Kohei Arai, editor, \emph{Intelligent Computing}, pages 565--576, Cham, 2023. Springer Nature Switzerland.
\newblock ISBN 978-3-031-37717-4.

\bibitem[Parashar et~al.(2022)Parashar, Kaur, Sharma, Singh, and Mishra]{Parashar2022revolutionary}
Binayak Parashar, Inderjeet Kaur, Anupama Sharma, Pratima Singh, and Deepti Mishra.
\newblock \emph{Revolutionary transformations in twentieth century: making AI-assisted software development}, pages 1--18.
\newblock De Gruyter, Berlin, Boston, 2022.
\newblock ISBN 9783110709247.
\newblock \doi{doi:10.1515/9783110709247-001}.
\newblock URL \url{https://doi.org/10.1515/9783110709247-001}.

\bibitem[Kitchenham et~al.(2009)Kitchenham, Pearl~Brereton, Budgen, Turner, Bailey, and Linkman]{kitchenham2009slr}
Barbara Kitchenham, O.~Pearl~Brereton, David Budgen, Mark Turner, John Bailey, and Stephen Linkman.
\newblock Systematic literature reviews in software engineering - a systematic literature review.
\newblock \emph{Inf. Softw. Technol.}, 51\penalty0 (1):\penalty0 7–15, jan 2009.
\newblock ISSN 0950-5849.
\newblock \doi{10.1016/j.infsof.2008.09.009}.
\newblock URL \url{https://doi.org/10.1016/j.infsof.2008.09.009}.

\bibitem[Brown et~al.(2020)Brown, Mann, Ryder, Subbiah, Kaplan, Dhariwal, Neelakantan, Shyam, Sastry, Askell, Agarwal, Herbert-Voss, Krueger, Henighan, Child, Ramesh, Ziegler, Wu, Winter, Hesse, Chen, Sigler, Litwin, Gray, Chess, Clark, Berner, McCandlish, Radford, Sutskever, and Amodei]{brown2020language}
Tom~B. Brown, Benjamin Mann, Nick Ryder, Melanie Subbiah, Jared Kaplan, Prafulla Dhariwal, Arvind Neelakantan, Pranav Shyam, Girish Sastry, Amanda Askell, Sandhini Agarwal, Ariel Herbert-Voss, Gretchen Krueger, Tom Henighan, Rewon Child, Aditya Ramesh, Daniel~M. Ziegler, Jeffrey Wu, Clemens Winter, Christopher Hesse, Mark Chen, Eric Sigler, Mateusz Litwin, Scott Gray, Benjamin Chess, Jack Clark, Christopher Berner, Sam McCandlish, Alec Radford, Ilya Sutskever, and Dario Amodei.
\newblock Language models are few-shot learners, 2020.

\bibitem[Necula(2023)]{necula2023artificial}
Sabina-Cristiana Necula.
\newblock Artificial intelligence impact on the labour force--searching for the analytical skills of the future software engineers.
\newblock \emph{arXiv preprint arXiv:2302.13229}, 2023.

\bibitem[Wu et~al.(2022)Wu, Gao, Zhang, Wang, and Tang]{wu2022bots}
Xiaojun Wu, Anze Gao, Yang Zhang, Tao Wang, and Yi~Tang.
\newblock A preliminary study of bots usage in open source community.
\newblock In \emph{Proceedings of the 13th Asia-Pacific Symposium on Internetware}, Internetware '22, page 175–180, New York, NY, USA, 2022. Association for Computing Machinery.
\newblock ISBN 9781450397803.
\newblock \doi{10.1145/3545258.3545284}.
\newblock URL \url{https://doi.org/10.1145/3545258.3545284}.

\bibitem[Wyrich et~al.(2021)Wyrich, Ghit, Haller, and M{\"u}ller]{wyrich2021bots}
Marvin Wyrich, Raoul Ghit, Tobias Haller, and Christian M{\"u}ller.
\newblock Bots don’t mind waiting, do they? comparing the interaction with automatically and manually created pull requests.
\newblock In \emph{2021 IEEE/ACM Third International Workshop on Bots in Software Engineering (BotSE)}, pages 6--10. IEEE, 2021.

\bibitem[Wessel et~al.(2022)Wessel, Gerosa, and Shihab]{wessel2022software}
Mairieli Wessel, Marco~A. Gerosa, and Emad Shihab.
\newblock Software bots in software engineering: Benefits and challenges.
\newblock In \emph{Proceedings of the 19th International Conference on Mining Software Repositories}, MSR '22, page 724–725, New York, NY, USA, 2022. Association for Computing Machinery.
\newblock ISBN 9781450393034.
\newblock \doi{10.1145/3524842.3528533}.
\newblock URL \url{https://doi.org/10.1145/3524842.3528533}.

\bibitem[Tony et~al.(2022)Tony, Balasubramanian, D\'{\i}az~Ferreyra, and Scandariato]{tony2022devbots}
Catherine Tony, Mohana Balasubramanian, Nicol\'{a}s~E. D\'{\i}az~Ferreyra, and Riccardo Scandariato.
\newblock Conversational devbots for secure programming: An empirical study on skf chatbot.
\newblock In \emph{Proceedings of the 26th International Conference on Evaluation and Assessment in Software Engineering}, EASE '22, page 276–281, New York, NY, USA, 2022. Association for Computing Machinery.
\newblock ISBN 9781450396134.
\newblock \doi{10.1145/3530019.3535307}.
\newblock URL \url{https://doi.org/10.1145/3530019.3535307}.

\bibitem[Savary-Leblanc et~al.(2023)Savary-Leblanc, Burgueño, Cabot, Le~Pallec, and Gérard]{santhanam2022bots}
Maxime Savary-Leblanc, Lola Burgueño, Jordi Cabot, Xavier Le~Pallec, and Sébastien Gérard.
\newblock Software assistants in software engineering: A systematic mapping study.
\newblock \emph{Software: Practice and Experience}, 53\penalty0 (3):\penalty0 856--892, 2023.
\newblock \doi{https://doi.org/10.1002/spe.3170}.
\newblock URL \url{https://onlinelibrary.wiley.com/doi/abs/10.1002/spe.3170}.

\bibitem[Pothukuchi et~al.(2023)Pothukuchi, Kota, and Mallikarjunaradhya]{pothukuchi2023impact}
Ameya~Shastri Pothukuchi, Lakshmi~Vasuda Kota, and Vinay Mallikarjunaradhya.
\newblock Impact of generative ai on the software development lifecycle (sdlc).
\newblock \emph{International Journal of Creative Research Thoughts}, 11\penalty0 (8), 2023.
\newblock URL \url{https://ssrn.com/abstract=4536700}.

\bibitem[Erlenhov et~al.(2020{\natexlab{b}})Erlenhov, de~Oliveira~Neto, Chukaleski, and Daknache]{erlenhov2020challenges}
Linda Erlenhov, Francisco~Gomes de~Oliveira~Neto, Martin Chukaleski, and Samer Daknache.
\newblock Challenges and guidelines on designing test cases for test bots.
\newblock In \emph{Proceedings of the IEEE/ACM 42nd International Conference on Software Engineering Workshops}, ICSEW'20, page 41–45, New York, NY, USA, 2020{\natexlab{b}}. Association for Computing Machinery.
\newblock ISBN 9781450379632.
\newblock \doi{10.1145/3387940.3391535}.
\newblock URL \url{https://doi.org/10.1145/3387940.3391535}.

\bibitem[Uzair and Naz(2023)]{uzair2023six}
Waqas Uzair and Sameen Naz.
\newblock Six-tier architecture for ai-generated software development: A large language models approach.
\newblock 2023.

\bibitem[Erlenhov et~al.(2022)Erlenhov, de~Oliveira~Neto, and Leitner]{erlenhov2022dependency}
Linda Erlenhov, Francisco~Gomes de~Oliveira~Neto, and Philipp Leitner.
\newblock Dependency management bots in open-source systems—prevalence and adoption.
\newblock \emph{PeerJ Computer Science}, 8, 2022.
\newblock \doi{10.7717/PEERJ-CS.849}.
\newblock URL \url{https://www.scopus.com/inward/record.uri?eid=2-s2.0-85128266032&doi=10.7717%2fPEERJ-CS.849&partnerID=40&md5=83062faf179c490146738deb92692235}.
\newblock Cited by: 2; All Open Access, Gold Open Access, Green Open Access.

\bibitem[Blocklove et~al.(2023)Blocklove, Garg, Karri, and Pearce]{blocklove2023}
Jason Blocklove, Siddharth Garg, Ramesh Karri, and Hammond Pearce.
\newblock Chip-chat: Challenges and opportunities in conversational hardware design.
\newblock \emph{CoRR}, abs/2305.13243, 2023.
\newblock \doi{10.48550/arXiv.2305.13243}.
\newblock URL \url{https://doi.org/10.48550/arXiv.2305.13243}.

\bibitem[Gupta et~al.(2022)Gupta, Epiphaniou, and Maple]{Gupta2022}
Sandeep Gupta, Gregory Epiphaniou, and Carsten Maple.
\newblock {AI-Augmented Usability Evaluation Framework for Software Requirements Specification}.
\newblock 6 2022.
\newblock \doi{10.36227/techrxiv.20097701.v1}.

\bibitem[Liu et~al.(2023)Liu, Yuan, Fu, Jiang, Hayashi, and Neubig]{Liu2023}
Pengfei Liu, Weizhe Yuan, Jinlan Fu, Zhengbao Jiang, Hiroaki Hayashi, and Graham Neubig.
\newblock Pre-train, prompt, and predict: A systematic survey of prompting methods in natural language processing.
\newblock \emph{ACM Comput. Surv.}, 55\penalty0 (9), jan 2023.
\newblock ISSN 0360-0300.
\newblock \doi{10.1145/3560815}.
\newblock URL \url{https://doi.org/10.1145/3560815}.

\bibitem[Yang et~al.(2023)Yang, Wang, Lu, Liu, Le, Zhou, and Chen]{yang2023large}
Chengrun Yang, Xuezhi Wang, Yifeng Lu, Hanxiao Liu, Quoc~V. Le, Denny Zhou, and Xinyun Chen.
\newblock Large language models as optimizers, 2023.
\newblock URL \url{https://doi.org/10.48550/arXiv.2309.03409}.

\bibitem[Chen et~al.(2023)Chen, Lin, Han, and Sun]{chen2023benchmarking}
Jiawei Chen, Hongyu Lin, Xianpei Han, and Le~Sun.
\newblock Benchmarking large language models in retrieval-augmented generation, 2023.
\newblock URL \url{https://arxiv.org/abs/2309.01431}.

\bibitem[Fielding(2000)]{fielding2000architectural}
Roy~Thomas Fielding.
\newblock \emph{Architectural styles and the design of network-based software architectures}.
\newblock University of California, Irvine, Ann Arbor, USA, 2000.

\end{thebibliography}

\end{document}